\documentclass[pra,twocolumn,showpacs,preprintnumbers,amsmath,amssymb]{revtex4}

\usepackage{graphicx}
\usepackage{dcolumn}
\usepackage{bm}

\def \ed {\end{document}}
\def\Fbox#1{\vskip1ex\hbox to 8.5cm{\hfil\fboxsep0.3cm\fbox{%
  \parbox{8.0cm}{#1}}\hfil}\vskip1ex\noindent}  

\def\be{\begin{equation}}\def\ee{\end{equation}}
\def\bea{\begin{eqnarray}}\def\eea{\end{eqnarray}}
\def\bse{\begin{subequations}}\def\ese{\end{subequations}}
\newcommand{\BE}[1]{\begin{equation}\label{#1}}
\newcommand{\BEA}[1]{\begin{eqnarray}\label{#1}}
\newcommand{\BSE}[1]{\begin{subequations}\label{#1}}

\let \= \equiv \let\*\cdot \let\~\widetilde \let\^\widehat \let\-\overline

\def\<{\left\langle}    \def\>{\right\rangle}
\def\({\left(}          \def\){\right)}
 \def \[ {\left [} \def \] {\right ]}

\begin{document}
\preprint{APS/123-QED}

\title{Synergy Dynamics of Vortices and Solitons in an Atomic Bose-Einstein Condensate Excited by an Oscillating Potential}

\author{Kazuya Fujimoto}
\affiliation{Department of Physics, Osaka City University, Sumiyoshi-ku, Osaka 558-8585, Japan}%

\author{Makoto Tsubota}
\affiliation{Department of Physics, Osaka City University, Sumiyoshi-ku, Osaka 558-8585, Japan}%

\date{\today}

\begin{abstract}
The hydrodynamics of quantized vortices and solitons in an atomic Bose-Einstein condensate excited by an oscillating potential are studied
by numerically solving the two-dimensional Gross-Pitaevskii equation. 
The oscillating potential keeps nucleating vortex dipoles, whose impulses alternatively change their direction synchronously with 
the oscillation of the potential. 
This leads to synergy dynamics of vortices and solitons which have not been previously reported in quantum fluids.
\end{abstract}

\pacs{67.85.De,03.75.Lm,67.25.dk,47.37.+q}%

\maketitle


\section{INTRODUCTION}

Topological defects are key concepts in general physics \cite{BunkovGodfrin}.
Quantum condensed systems are ideal for studying topological defects. 
Quantized vortices and solitons have been thoroughly characterized in superfluid $^4$He \cite{Donnelly} and $^3$He \cite{Vollhardt,Volovik}. 
Modern research developments on quantum turbulence (QT) is still based on understanding dynamics of quantized vortices \cite{PLTP}. 
An atomic Bose-Einstein condensate (BEC) can be used to investigate quantum hydrodynamics. 
Atomic BECs have several advantages over superfluid helium. In particular, modern optical techniques enable control of the condensate and direct visualization of topological defects such as vortices and solitons. 
Actually many important works have been performed on quantized vortices these years \cite{FetterRMP}. 
Recently Henn {\it et al.} made and observed QT in a BEC by introducing an external oscillatory perturbation of the trapping potential \cite{Henn09}.

Here we numerically address the response of a BEC to an oscillating repulsive potential.
Vibrating structures such as spheres, grids, and wires are used in superfluid $^4$He and $^3$He to create QT \cite{PLTP, Hanninen07}.  
Despite the differences between these structures, the experiments show surprisingly similar behavior.
We apply this methodology for atomic BECs for the first time. 

A few works on oscillating potentials in atomic BECs have already been numerically and experimentally performed \cite{Jackson00, Raman99, Onofrio20}. 
Dissipation which works above some critical velocity was investigated theoretically by Jackson {\it et al.} \cite{Jackson00} and 
experimentally by Raman {\it et al.} \cite{Raman99}. 
However, they did not observe the dynamics of vortices and solitons reported in this work. 
Recently, Neely {\it et al.} observed the formation and dynamics of vortex dipoles by forcing superflow around a repulsive Gaussian potential within an oblate BEC \cite{Neely10}. 

Studying quantum hydrodynamics in an atomic BEC subject to an oscillating potential should open up a research area different from helium and other 
BEC cases for the following reasons.
First, since an atomic BEC is a clean system free of remnant defects and impurities, it is possible to study the intrinsic nucleation of topological defects. 
Second, the oscillating potential leads to synergy dynamics of quantized vortices and solitons that have not been previously observed. Linear motion of the 
potential results in vortex dipoles with {\it fixed} charge \cite{Neely10}, 
whereas oscillatory motion leads to vortex dipoles with {\it alternating} charges, causing rearrangement of dipoles and a metamorphosis between vortices and solitons. 
Third, the oscillation introduces another important parameter, namely {\it frequency}, into quantum hydrodynamics. Nucleation of defects and the transition to QT depend on frequency, which should be investigated in future. 
Fourth, this work develops a powerful new method for making QT in a trapped BEC in addition to other known ones \cite{Henn09,Berloff02, Kobayashi07}.
Eventually the dynamics of vortices and solitons can be visualized in atomic BECs, enabling a direct comparison between experiments and theoretical or numerical results. 

In this work we investigate the quantum hydrodynamics of a trapped BEC with an oscillating potential by numerically solving the two-dimensional Gross-Pitaevskii (GP) equation. In Section II the model and numerical calculation are described. 
Section III presents the whole dynamics of vortices obtained by our numerical calculation. 
Next we explain elementary processes related to vortices and solitons in Section IV. 
The oscillating potential may heat the condensate, which is estimated in Section V. 
Finally we summarize our work in Section VI.

\begin{figure*}[!t]
\begin{center}
\includegraphics[keepaspectratio, width=15cm]{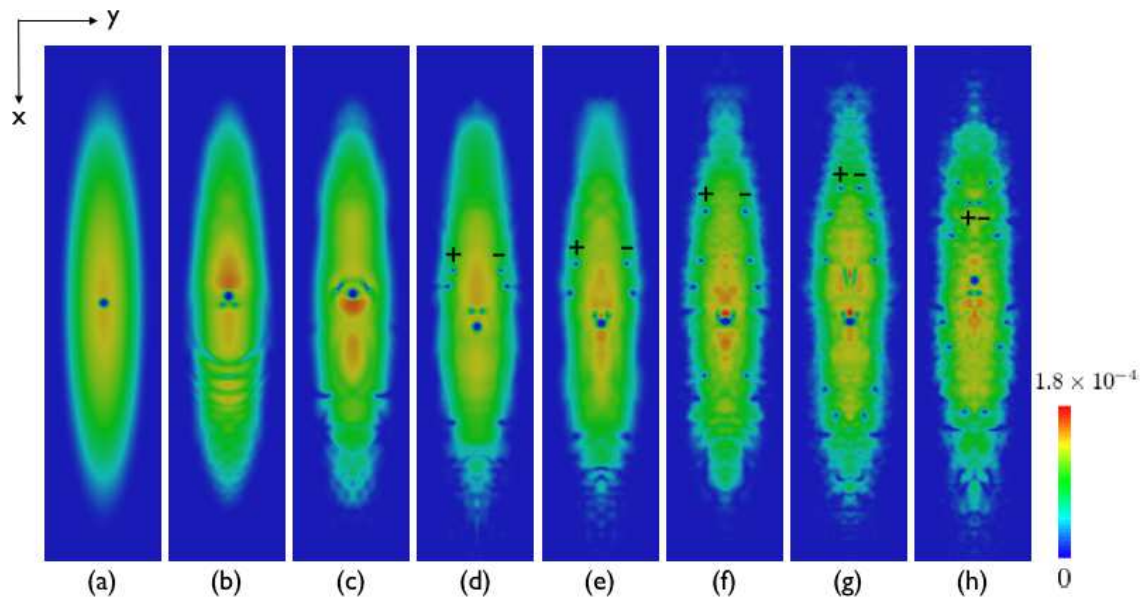}
\end{center}
\caption{(color online) Migration of vortices. Density profile at (a)\hspace{0.5mm}$t$=0\hspace{0.5mm}s, (b)\hspace{0.5mm}$t$=0.0345\hspace{0.5mm}s, (c)\hspace{0.5mm}$t$=0.0591\hspace{0.5mm}s, (d)\hspace{0.5mm}$t$=0.0760\hspace{0.5mm}s, (e)\hspace{0.5mm}$t$=0.0845\hspace{0.5mm}s, (f)\hspace{0.5mm}$t$=0.149\hspace{0.5mm}s, (g)\hspace{0.5mm}$t$=0.211\hspace{0.5mm}s, and 
(h)\hspace{0.5mm}$t$=0.240\hspace{0.5mm}s. The $x$ and $y$ dimensions of the images are $140\hspace{0.5mm}\rm 
{\mu m}$ and $32.5\hspace{0.5mm}\rm {\mu m}$. 
The symbol + (-) denotes a vortex with clockwise (counter-clockwise) circulation. 
There are two kinds of holes. 
The central large hole is the Gaussian potential, while the other small holes are vortices.}
\end{figure*}

\section{The MODEL AND NUMERICAL CALCULATION}
We consider a dilute atomic BEC, assuming that the condensate is pancake shaped. This system is well described by 
a macroscopic wavefunction $\psi$ obeying the GP equation 
\begin{equation}
i\hbar\frac{\partial \psi}{\partial t}
=-\frac{\hbar^2}{2m}\nabla^2 \psi+V\psi
+g \vert \psi \vert^2 \psi ,
\end{equation}
where $m$ is the particle mass, $V$ is the potential, and $g$ is an interaction parameter for the two-dimensional case. The wavefunction 
$\psi$ is normalized by the total particle number $N$. Suppose the condensate is confined by a harmonic potential 
$V_{\rm h}$ and penetrated by a Gaussian potential $V_{\rm G}$, so that $V=V_{\rm h}+V_{\rm G}$ where $V_{\rm h}=\frac{1}{2}
m(\omega _{x}^2x^2+\omega _{y}^2y^2)$ and $V_{\rm G} = V_0{\rm exp}[-((x-x_0(t))^2+y^2)/d^2]$. Here $x_{0}(t)$ is the $x$-coordinate of the center of the Gaussian potential and $d$ is its radius. Oscillate the Gaussian potential as $x_{0}(t)
=\epsilon \hspace{0.5mm} \rm{sin}(\omega \it{t})$. 
We use a dimensionless form of Eq.\hspace{0.5mm}(1) for numerical calculations. 
Space and time are normalized by $\hbar/ \sqrt{2mgn_0}$ and $\hbar/gn_0$, where $n_0$ is the density near the center of the condensate. Choose parameters 
$ g=4.19 \times 10^{-45} \hspace{0.5mm}\rm{J/m^2} $, $ m=1.42 \times 10^{-25} \hspace{0.5mm}\rm{kg} $, $ N=6.6 \times 10^{4} $, $ \omega _x= 2 \pi \times 5 $\hspace{0.5mm}/s, 
$ \omega _y= 2 \pi \times 25 $\hspace{0.5mm}/s, $d=0.6 \hspace{0.5mm}\mu \rm m$, $\epsilon =7 \hspace{0.5mm}\rm{\mu} m$, $\omega=100$\hspace{0.5mm}/s and $ V_0= 60gn_0 $. 
We use Crank-Nicholson method to perform numerical calculations without dissipation and noise.
For the simulation, space in the $x$ and $y$ directions is discretized into 2048$\times$640 bins.

\section{WHOLE DYNAMICS}
The dynamics of vortices in which they experience a lengthy migration are shown in Fig.\hspace{0.5mm}1. 
Following the destiny of vortices nucleated by the oscillating potential enables us to survey their dynamics. 
The initial state in the static Gaussian potential in Fig.\hspace{0.5mm}1(a) is obtained by an imaginary time step of the GP equation. 
A vortex pair is nucleated behind the Gaussian potential in Fig.\hspace{0.5mm}1(b) as the potential starts to move. 
Then the oscillating potential nucleates vortex pairs whose impulses alternately change direction.
They reconnect with each other to make new vortex pairs, leaving the potential in Fig.\hspace{0.5mm}1(c). 
This phenomenon is not observed for the case of uniform motion of the potential, but only for an oscillating potential. 
Reaching the surface in Fig.\hspace{0.5mm}1(d), the vortex pairs interact with ghost vortices, which are vortices in the low-density 
region and described later in IV.A. Then the vortices head toward the 
bow of the condensate along the surface in Fig.\hspace{0.5mm}1(e) and (f). 
A vortex coming up from the left side reaches the bow to meet one from the right side in Fig.\hspace{0.5mm}1(g), thus making a new vortex pair.
Finally, the pair comes back to the center of the condensate in Fig.\hspace{0.5mm}1(h). 
Thus the vortices nucleated by the potential enjoy a lengthly migration in the "sea" of BEC; the vortices are nucleated from the potential, reconnect, move away from it, reach the surface, head toward the bow and come back to the center. 
In the following, we illustrate elementary processes related to the synergy dynamics.

\begin{figure}[!t]
\begin{center}
\includegraphics[keepaspectratio, width=8cm]{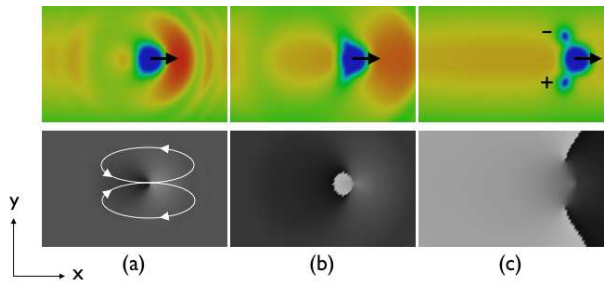}
\end{center}
\caption{(color online) Vortex nucleation by the potential: (a) $t$=0.00253\hspace{0.5mm}s, (b) $t$=0.00490\hspace{0.5mm}s, and (c) $t$=0.0118\hspace{0.5mm}s. 
The upper and lower panels are density and phase profiles. 
The $x$ and $y$ dimensions of each image are $13 \hspace{0.5mm}\rm{\mu m}$ and $8\hspace{0.5mm}\rm{\mu m}$.
The symbol + (-) denotes a vortex with clockwise (counter-clockwise) circulation. 
Black arrows in the density profile indicate the direction of motion of the potential, 
while the white arrows in the phase profile show the backflow. The value of the phase varies from -$\pi$ (white) to $\pi$ (black). }
\end{figure}

\section{ElEMENTARY PROCESSES}
In this section we discuss elementary processes, which are nucleation of vortices, reconnection of vortex pairs near the potential, divorce of vortex 
pair at the surface, nucleation of solitons and collapse of solitons. These processes occur in the dynamics of Fig.\hspace{0.5mm}1.

\subsection{Nucleation of vortices}
A key issue is to understand how quantized vortices are nucleated. 
For a condensate to have vortices, it should bring some seeds of the topological defects from somewhere. 
This situation is reminiscent of the formation of a vortex lattice in a rotating BEC \cite{Tsubota02}. Rotation nucleates "ghost vortices" 
in the low-density region at the outskirts of the condensate, and they are dragged into the interior to become the usual vortices having the condensate density.
Our condensate has to do a similar thing. 
Ghost vortices can appear in two low-density regions in our system; one is inside the oscillating potential and the other is at the outskirts of the condensate.
As soon as the potential starts to oscillate, a pair of ghost vortices is nucleated inside the potential and the condensate surface is filled with them. 

Vortex nucleation induced by the potential is shown in Fig.\hspace{0.5mm}2. 
It follows that ghost vortices inside the potential cause vortex nucleation. 
We consider nucleation of $both$ kinds of vortices in this work. There are two critical velocities; $v_{\rm c1}$ for a pair of ghost vortices and $v_{\rm c2}$ 
for a pair of vortices with $ v_{\rm c1}<v_{\rm c2} $. When the velocity $v_{\rm p}=\epsilon \hspace{0.5mm} \omega$ of the potential is smaller than $v_{\rm c1}$, vortices are not nucleated in 
the condensate. When $v_{\rm p}$ exceeds $v_{\rm c1}$, a pair of ghost vortices is nucleated inside the Gaussian potential. If $v_{\rm p}$ is smaller than 
$v_{\rm c2}$, however, those ghost vortices are annihilated. 
The repeated nucleation and annihilation of pairs of ghost vortices are characteristic 
of the oscillating case with $v_{\rm c1} < v_{\rm p} < v_{\rm c2}$. 
The observable critical velocity is $v_{\rm c2}\sim 580\hspace{0.5mm} \rm{\mu m/s}$ for the present case, but $v_{\rm c2}$ 
strongly depends on the frequency $\omega$ of the oscillation, which is an important target in future. 
The critical velocity $v_{\rm c2}$ is smaller than the sound speed $1.33
\times 10^{3}$\hspace{0.5mm}$\rm \mu m/s$, which is reported by Jackson {\it et al.}\hspace{0.5mm}\cite{Jackson00}. 
When $v_{\rm p}$ exceeds $v_{\rm c2}$, vortices are nucleated. 
These processes are shown in Fig.\hspace{0.5mm}2. A velocity field with backflow is induced by the potential in Fig.\hspace{0.5mm}2(a). Then a pair 
of ghost vortices is nucleated inside the potential in Fig.\hspace{0.5mm}2(b). Finally, the ghost vortices move away from it and a vortex pair appears 
in Fig.\hspace{0.5mm}2(c).

\begin{figure}[!t]
\begin{center}
\includegraphics[keepaspectratio, width=7cm,clip]{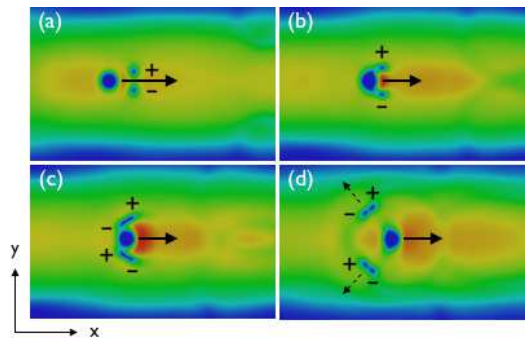}
\end{center}
\caption{(color online) Reconnection of vortices near the potential: Density profile at (a)\hspace{0.5mm}$t$=0.0507\hspace{0.5mm}s, (b)\hspace{0.5mm}$t$=0.0551\hspace{0.5mm}s, (c)\hspace{0.5mm}$t$=0.0571\hspace{0.5mm}s, (d)\hspace{0.5mm}$t$=0.0596\hspace{0.5mm}s. 
The $x$ and $y$ dimensions of each image are $39\hspace{0.5mm}\rm {\mu m}$ and $24\hspace{0.5mm}\rm {\mu m}$. 
The symbol + (-) denotes a vortex with clockwise (counter-clockwise) circulation. 
Black arrows indicate the direction of motion of the potential. 
The dashed arrows indicate the direction of the motion for vortices in (d).}
\end{figure}

\begin{figure*}[!t]
\begin{center}
\includegraphics[keepaspectratio, width=14cm,clip]{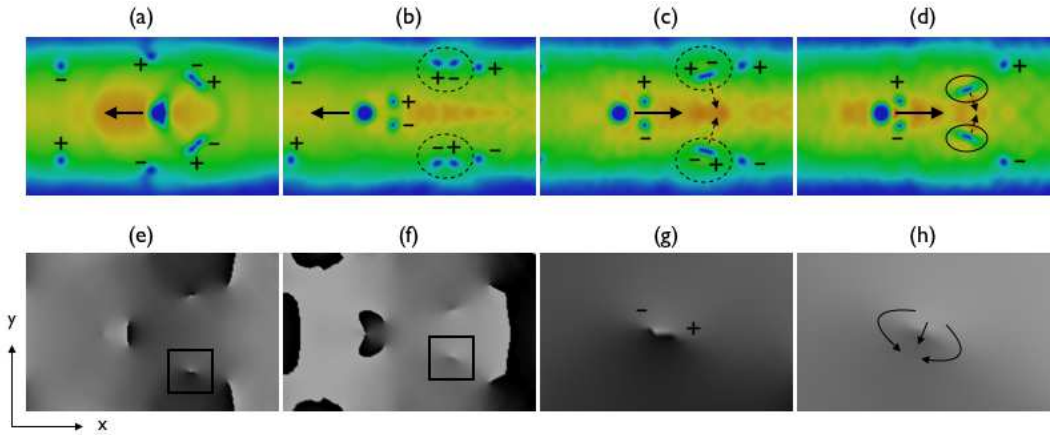}
\end{center}
\caption{(color online) Nucleation of solitons: (a)\hspace{0.5mm}$t$=0.0929\hspace{0.5mm}s, (b)\hspace{0.5mm}$t$=0.106\hspace{0.5mm}s, (c)\hspace{0.5mm}$t$=0.111\hspace{0.5mm}s, (d)\hspace{0.5mm}$t$=0.114\hspace{0.5mm}s. 
The panels (a)$\sim$(d) and (e)$\sim$(h) show the density(upper) and phase(lower) profile. 
The $x$ and $y$ dimensions of (a)$\sim$(f) and (g), (h) are $39\hspace{0.5mm}\rm {\mu m} \times 24\hspace{0.5mm}\rm {\mu m}$ 
and $9\hspace{0.5mm}\rm {\mu m} \times 6\hspace{0.5mm}\rm {\mu m}$.
The phase profiles for (c) and (d) are respectively (e) and (f).  
The figures (g) and (h) are the enlarged figures corresponding to the square boxes in (e) and (f). 
The symbol + (-) denotes a vortex with clockwise (counter-clockwise) circulation. 
Black arrows indicate the direction of motion of the potential in (a)$\sim$(d), but it shows the velocity field in (h). 
In the process from (a) to (b), reconnection of vortices occurs, then two new vortex pairs, which will become solitons, are created. 
These pairs are depicted by closed loops with dashed lines. The closed loops with solid lines show the solitons. 
The dashed arrows indicate the direction of the motion for vortices in (c), and one for solitons in (d). The value of the phase varies from -$\pi$ (white) to $\pi$ (black).}
\end{figure*}

\begin{figure}[!b]
\begin{center}
\includegraphics[keepaspectratio, width=7cm,clip]{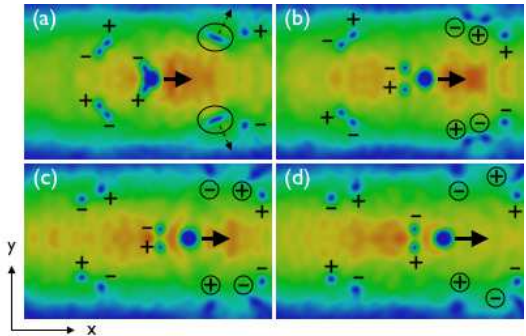}
\end{center}
\caption{(color online) Collapse of solitons: Density profile at (a)\hspace{0.5mm}$t$=0.126\hspace{0.5mm}s, (b)\hspace{0.5mm}$t$=0.131\hspace{0.5mm}s, 
(c)\hspace{0.5mm}$t$=0.136\hspace{0.5mm}s, (d)\hspace{0.5mm}$t$=0.138\hspace{0.5mm}s. 
The $x$ and $y$ dimensions of each image are $39\hspace{0.5mm}\rm {\mu m}$ and $24\hspace{0.5mm}\rm {\mu m}$. 
The symbol + (-) denotes a vortex with clockwise (counter-clockwise) circulation. 
Black arrows indicate the direction of motion of the potential. The closed loops with solid lines show the solitons. 
The symbols $\oplus$ and $\ominus$ from (b) to (d) denote the vortices created through the collapse of the soliton. 
The dashed arrows indicate the direction of the motion for solitons in (a).}
\end{figure}

\subsection{Reconnection of vortex pairs near the potential}
Figure\hspace{0.5mm}1(c) shows a phenomenon characterized by the oscillating case after vortex nucleation,
as detailed in Fig.\hspace{0.5mm}3(a) to (d). 
After a vortex pair is nucleated behind the potential, it has an impulse in the direction of the velocity of the potential.  
Hence the pair follows the potential in Fig.\hspace{0.5mm}3(a). Thereafter the oscillating potential changes the direction of its velocity 
in Fig.\hspace{0.5mm}3(b) and it makes a new ghost vortex pair whose direction of impulse is opposite to that of the old pair. 
Then the ghost pair gets away from the potential, becoming a new vortex pair.  
The pair immediately reconnects with the old pair in Fig.\hspace{0.5mm}3(c). 
This process produces new vortex pairs leaving the potential, and the pairs go toward the surface of the condensate in Fig.\hspace{0.5mm}3(d).

\subsection{Divorce of vortex pairs at the surface}
The vortex pair escaping from the potential in Fig.\hspace{0.5mm}3(d) reaches the surface and 
interacts with many fluctuating ghost vortices there. The pair divorces, and the two resulting vortices move in opposite directions along the surface.
The motion is caused by the following mechanism. 
The normal component of the velocity field at the surface is suppressed when vortices are close to the surface. 
This makes the velocity field parallel to the surface, which moves the vortices along the surface. 
This motion may be understood shortly by using the idea of image vortices \cite{image}, though 
the idea assumes that the boundary is a solid wall and this is not exactly the case.

\subsection{Nucleation of solitons}
The most interesting phenomenon in our research is the synergy dynamics between solitons and vortices. 
Nucleation, collision, and collapse of solitons are observed. 
These occur after Fig.\hspace{0.5mm}1(e). 
A soliton can be identified by its low density and phase jump of about $\pi$.
Figure\hspace{0.5mm}4(a) through (h) show the nucleation of a soliton, which is peculiar to the dynamics by an oscillating potential. After the process 
in Fig.\hspace{0.5mm}3(a) to (d) occurs twice in both the right and left regions, the configuration of Fig.\hspace{0.5mm}4(a) is obtained. 
The vortex pairs in the right region in Fig.\hspace{0.5mm}4(a) get near the surface, and reconnect 
with other vortices in Fig.\hspace{0.5mm}4(b). The vortex pairs are depicted by closed loops with dashed lines. The pairs head toward the center of 
the condensate in Fig.\hspace{0.5mm}4(c). Subsequently, annihilation of a pair occurs as it reaches the center, becoming a soliton in Fig.\hspace{0.5mm}4(d). 
The change from a vortex dipole to a soliton is related to the nucleation of rarefaction pulse \cite{Robert1982, Natalia02}.

One may think that the edges of the solitons have quantized vortices so that the low density regions inside the closed loops in Fig.\hspace{0.5mm}4(d) are merely vortex dipoles. 
However, there is a clear distinction between a vortex dipole and a soliton, which is obviously seen in the phase profiles (Fig.\hspace{0.5mm}4(e), (h)) corresponding to Fig.\hspace{0.5mm}4(c) and (d). Figure\hspace{0.5mm}4(e) and (f) show the phase profiles before and after the annihilation of the dipoles. 
Their enlarged phase profiles are Fig.\hspace{0.5mm}4(g) and (h), which show respectively the vortex dipole and the soliton. 
Figure 4(g) shows apparently that the dipole has the topological defects in the phase profile, while the soliton in Fig.\hspace{0.5mm}4(h) 
does not have such defects and the velocity field around the edges does not have quantized circulation as seen by the arrows referring to superflow.  
There remains the velocity field related to quantized vortices outside the edges, whereas the velocity field which tends to cancel the outside flow 
are induced in the low density region between the edges. Thus we can identify the soliton by checking the density and phase profile.

\subsection{Collapse of solitons}
The collapse of the solitons nucleates vortex pairs in (a) to (d) of Fig.\hspace{0.5mm}5. 
The solitons in Fig.\hspace{0.5mm}4(d) have opposite impulse, so that they collide with each other. 
However, they pass through each other without changing shape, in accord with the nature of solitons. 
The solitons then move toward the surface in Fig.\hspace{0.5mm}5(a). A two-dimensional soliton is unstable, and collapse of the solitons occurs at the surface. 
As a result, a soliton decays into a vortex pair at the surface, as shown in Fig.\hspace{0.5mm}5(b). 
These vortex pairs are depicted by closed loops with solid lines. 
Immediately after the collapse, the vortex pairs interact with ghost vortices in Fig.\hspace{0.5mm}5(c). 
Then, the vortices move toward the bow of the condensate along the surface in Fig.\hspace{0.5mm}5(d) because of the mechanism described in IV.C. 
This process repeats as long as the Gaussian potential oscillates.
This sort of transformation between vortex dipoles and solitons has been reported by Huang {\it et al.} \cite{Huang23}.

\section{HEATING OF THE CONDENSATE}
It is possible to heat a condensate by an oscillating potential. 
We numerically calculate the increase in the total energy, estimating the temperature change 
by using the specific heat of the equilibrium state of the system. The change is found to be only $1$\hspace{0.5mm}nK even by the time of the
last image in Fig.\hspace{0.5mm}1. Hence the heating is negligible, and use of the GP model is valid.

\section{CONCLUSION}
We have performed numerical calculations of the two-dimensional GP equation to investigate the dynamics of vortices and solitons in a trapped BEC induced 
by an oscillating potential. 
This paper reveals the dynamics characterized by oscillation. It is essential that the oscillating potential makes vortex dipoles with different 
charges, which results in the synergy dynamics between vortices and solitons.  
The parameters used in these calculations are appropriate to experiments on atomic BECs. 
We confirm the dynamics not only for the single fixed geometry but also for similar sets of parameters and geometry. 
Hence, the dynamics obtained in our study should be experimentally observable. 
However, the dynamics by the oscillating potential should be dependent on the frequency and amplitude of the oscillation. 
Preliminary calculations indicate that a potential with 
lower amplitude creates more solitons, in contrast to the present result. In addition, the aspect ratio of the condensate shape should affect the dynamics. 
The systematic studies of how the dynamics depends on the parameters or geometry are reported elsewhere. 

\section*{ACKNOWLEDGMENT}
The authors are grateful to H. Takeuchi for useful discussions. M. T. acknowledges the support of a Grant-in-Aid for Science Research from JSPS (Grant No.21340104).



 
\end{document}